\documentclass[10pt,aps,prl,twocolumn,showpacs,amsmath,amssymb,groupedaddress]{revtex4-1}
\usepackage{graphicx}
\begin{document}
\title{Searching for Supersolidity in Ultracold Atomic Bose Condensates with Rashba Spin-Orbit Coupling  }
\author{Renyuan Liao}
\affiliation{Fujian Provincial Key Laboratory for Quantum Manipulation and New Energy Materials, College of Physics and Energy, Fujian Normal University, Fuzhou 350117, China}
\affiliation{Fujian Provincial Collaborative Innovation Center for Optoelectronic Semiconductors and Efficient Devices, Xiamen, 361005, China}
\date{\today}
\begin{abstract}
   We developed functional integral formulation for the stripe phase of a spinor Bose-Einstein condensates with Rashba spin-orbit coupling. The excitation spectrum is found to exhibit double gapless band structures, identified to be two Goldstone modes resulting from spontaneously broken internal gauge symmetry and translational invariance symmetry. The sound velocities display anisotropic behaviors with the lower branch vanishes in the direction perpendicular to the stripe in the x-y plane. At the transition point between the plane wave phase and the stripe phase, physical quantities such as fluctuation correction to the ground state energy and quantum depletion of the condensates exhibit discontinuity, characteristic of the first order phase transition. Despite strong quantum fluctuations induced by Rashba spin-orbit coupling, we show that the supersolid phase is stable against quantum depletion. Finally we extend our formulation to finite temperatures to account for interactions between excitations.
\end{abstract}
\pacs{67.85.Fg, 03.75.Mn, 05.30.Jp, 67.85.Jk}
\maketitle

One of the most intriguing predictions of the quantum theory is the possibility of supersolidity. It combines superfluid flow with long-range spatial periodicity of solids, two properties that are often mutually exclusive. This quantum phase requires the breaking of two continuous symmetries: the phase invariance of the superfluid and the continuous translational invariance to form the crystal~\cite{BON12,CHA13}. It has been conjectured since 1970 that it might be possible to create a supersolid in solid $^4$He~\cite{CHE70}. Despite considerable theoretical and experimental efforts~\cite{CHA04,DAY07,CHA12,KUK14}, the quest for supersolidity in solid $^4$He remains elusive.

Over the past decade, physicists have pursued an alternate route to supersolidity, using the rapidly developing techniques for engineering ultracold quantum matter~\cite{DAL11,RIT13,GOL14}. There have been systematic theoretical efforts for realizing supersolidity in atomic gases with dipolar~\cite{PUP10,POL10} and soft core, finite range interactions~\cite{CIN10,SAC12,KUN12}. In 2017, two research groups from ETH Zurich~\cite{ETH17} and from MIT~\cite{MIT17} reported on the first creation of a supersolid with ultracold quantum gases. The Zurich group prepared Bose-Einstein condensates inside two optical cavities, which enhanced atomic interactions until they started to spontaneously crystallize and form a solid that maintains the inherent superfluidity of the Bose-Einstein condensates. The MIT group created effective one-dimensional spin-orbit coupling (SOC) to realize a stripe phase with density modulations verified through Bragg reflection.

So far, spinor Bose-Einstein condensates with one-dimensional SOC~\cite{LIN11,ZHA12} and two-dimensional SOC~\cite{ZHA16} have been realized in experiments. On the theoretical side, a great deal of attention has been paid to the ground-state phase diagram and properties of the plane-wave phase~\cite{HO11,ZHA10,YUN12,BAY12,BAR12,XU12,CUI13,LIA13}. Apart from a beautiful work on the study of the stripe phase with one-dimensional SOC~\cite{YUN13}, exploration of the stripe phase with two-dimensional SOC is still lacking.
In anticipation of immediate experimental relevance, investigation of  supersolid phase in Bose-Einstein condensates subject to Rashba SOC has become an interesting and urgent task. In this work, we shall fill in this gap by carrying out a comprehensive study.

We consider three-dimensional homogeneous two-species Bose gases with an isotropic in-plane Rashba spin-orbit coupling, described by the following grand canonical Hamiltonian,
\begin{eqnarray}
 H&=&\int d^3\mathbf{r}\sum_{\sigma=\uparrow,\downarrow}\left[\psi_\sigma^\dagger\left(-\frac{\hbar^2\nabla^2}{2m}-\mu\right)\psi_\sigma+g(\psi_\sigma^\dagger\psi_\sigma)^2\right]\nonumber\\
 &&+\int d^3\mathbf{r}\left[2g_{\uparrow\downarrow}\psi_\uparrow^\dagger\psi_\uparrow\psi_\downarrow^\dagger\psi_\downarrow+ (\psi_\uparrow^\dagger \hat{R}\psi_\downarrow+h.c.)\right],
 \label{eq1}
\end{eqnarray}
where bosonic operators $\psi_\sigma^\dagger$ and $\psi_\sigma$ satisfy commutation relation $[\psi_\sigma(\mathbf{r}),\psi_{\sigma^\prime}^\dagger(\mathbf{r^\prime})]=\delta_{\sigma\sigma^\prime}\delta^3(\mathbf{r-r^\prime})$, $\mu$ is the chemical potential, $g$ and $g_{\uparrow\downarrow}$ characterizes intra-species and inter-species interaction, respectively, and the in-plane spin-orbit coupling is described by $\hat{R}=\lambda(\hat{p_x}-i\hat{p_y})$, with $\lambda$ being the coupling strength. For brevity, we shall take natural units by setting $\hbar=2m=k_B=1$ from now on. We choose $gn_0$ as the basic energy scale, then the corresponding momentum scale is $\sqrt{gn_0}$, with $n_0$ being the density of the condensates.

Remarkable properties of this Hamiltonian stem from its translational invariance. For a non-interacting system, the lowest energy states are huge degenerate, lying at the circle of ``Rashba-ring" defined by the equation $q_x^2+q_y^2=(\lambda/2)^2$. When interactions is turned on, the system may favor a plane-wave phase or a striped phase, depending on whether $g>g_{\uparrow\downarrow}$ or  $g<g_{\uparrow\downarrow}$. The plane wave phase is characterized by a single condensation momentum, while the stripe phase involves linear combinations of pairs of plane-waves with opposite momenta, spontaneously breaking translation invariance. From general arguments one expects that the spontaneously breaking of this continuous symmetry is at the origin of a new gapless Goldstone mode. Without loss of generality, we shall assume that  the condensation momentum lies along x axis with $\mathbf{K}=K\hat{x}$. We take the form of the condensate wave function for pseudo-spin $\sigma$ as $\phi_{0\sigma}(\mathbf{r})=\sum_\alpha\phi_{0\alpha\sigma}e^{i(2\alpha-1)\mathbf{K}\cdot\mathbf{r}}$, where $\alpha$'s are integers.

\begin{figure}[t]
\includegraphics[width=1\columnwidth,height=0.6\columnwidth]{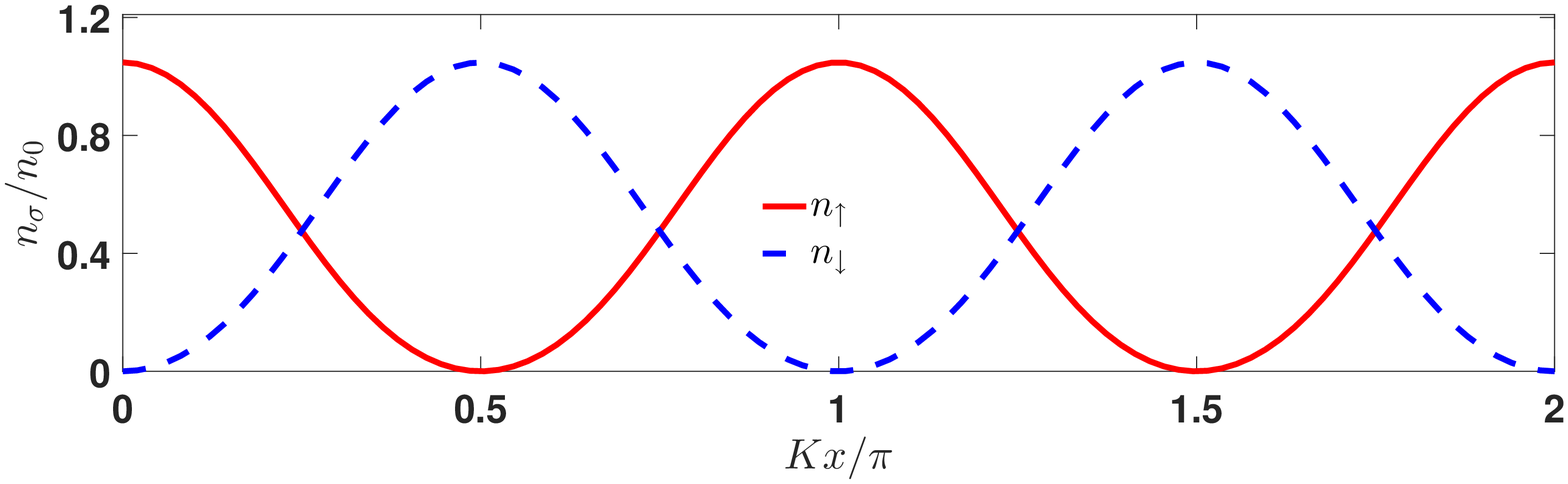}
\caption{(color online). Ground-state density profiles $n_\uparrow/n_0$ and $n_\downarrow/n_0$ in the x direction where translational invariance is broken. The density modulation has a spatial period $d=\pi/K$ with $K/\sqrt{gn_0}=0.9905$, corresponding to a reciprocal wave vector $2\mathbf{K}$. The parameters used here are: $\lambda/\sqrt{gn_0}=2$ and $g_{\uparrow\downarrow}/g=2$. The chemical potential can be determined via $\partial\Omega_0/\partial n_0=0$. }
\label{fig1}
\end{figure}

By minimizing the ground state energy with respect to variational parameters $\phi_{0\alpha\sigma}$ and $K$ with the constraint that the total number density of particle is fixed as $n_0$, we obtain mean-field configuration of the condensate wave function, involving terms with opposite phase ($e^{\pm iKx}$, $e^{\pm i3Kx}$, ...), responsible for the density modulations. The resultant ground state density profile is shown in Fig.~\ref{fig1}. It is evident that the density distribution shows periodic modulation with a spatial period $d=\pi/K$, indicating the existence of a diagonal long-range order.

We proceed to consider the effects of quantum fluctuations on top of the mean-field ground state. To treat the problem in a systematic way, we resort to the formulation of functional integral. The partition function of the system can be casted as $\mathcal{Z}=\int d[\psi_\sigma^\dagger,\psi_\sigma]e^{-S}$ with the action given by~\cite{SIM10} $S=\int_0^{1/T} d\tau \left[\int d^3\mathbf{r}\sum_\sigma \psi_\sigma^\dagger\partial_\tau \psi_\sigma+H(\psi_\sigma^\dagger,\psi_\sigma)\right]$, where $T$ is the temperature. In the spirit of Bogoliubov theory, we separate the Bose field $\psi_\sigma$ into a mean-field part $\phi_{0\sigma}$ and a fluctuating part $\phi_\sigma$ as $\hat{\psi}_\sigma=\phi_{0\sigma}+\hat{\phi}_{\sigma}$, and keep the action up to the quadratic order, yielding an effective action $S_{eff}=S_0+S_g=\int d\tau d^3\mathbf{r} (\mathcal{L}_0+\mathcal{L}_g)$ with
\begin{subequations}
\begin{eqnarray}
 \mathcal{L}_0&=&\sum_{\sigma}\left[\phi_{0\sigma}^*(-\nabla^2-\mu)\phi_{0\sigma}+gn_{0\sigma}^2\right]\nonumber\\
 &&+2g_{\uparrow\downarrow}n_{0\uparrow}n_{0\downarrow}+\left(\phi_{0\uparrow}^*\hat{R}\phi_{0\downarrow}+c.c\right),\\
 \mathcal{L}_g&=&\sum_\sigma\left[\phi_{\sigma}^\dagger(\partial_\tau+\hat{\xi}_\sigma)\phi_{\sigma}+g(\phi_{0\sigma}^2\phi_{\sigma}^{\dagger 2}+h.c.)\right]+(\phi_\uparrow^*\hat{R}\phi_\downarrow+h.c.)\nonumber\\
 &+&2g_{\uparrow\downarrow}\left[(\phi_{0\uparrow}^*\phi_{0\downarrow}^*\phi_\uparrow\phi_\downarrow+\phi_{0\uparrow}^*\phi_{0\downarrow}\phi_\uparrow\phi_\downarrow^\dagger)+h.c.\right].
\end{eqnarray}
\end{subequations}
In the above, we have used shorthand notations: $n_{0\sigma}=\phi_{0\sigma}^*\phi_{0\sigma}$ and $\hat{\xi}_\sigma=-\nabla^2-\mu+4gn_{0\sigma}+2g_{\uparrow\downarrow}n_{0\bar{\sigma}}$.
Up to this level, it is a formal manipulation of the gaussian action  $S_g$ on top of the mean-field action $S_0$. Guided by the form of mean-field solution $\phi_{0\sigma}(\mathbf{r})$, we expand the fluctuating fields $\hat{\phi}_\sigma(\mathbf{r})$ as follows
\begin{eqnarray}
  \begin{pmatrix}
     \phi_\uparrow(\mathbf{r}) \\ \phi_\downarrow(\mathbf{r}) \end{pmatrix}
 =\sum_{\mathbf{q}\alpha}\begin{pmatrix}\phi_{\mathbf{q}\alpha\uparrow}\\ \phi_{\mathbf{q}\alpha\downarrow} \end{pmatrix}e^{i(2\alpha-1)\mathbf{K}\cdot\mathbf{r}}e^{i\mathbf{q}\cdot\mathbf{r}}.
 \label{eq3}
\end{eqnarray}
Substituting Eq.~($\ref{eq3}$) into Eq.~(2b), we obtain
\begin{widetext}
\begin{eqnarray}
   S_g&=&\sum_{\mathbf{q}\alpha\sigma}\phi_{\mathbf{q}\alpha\sigma}^\dagger[-iw_n+\mathbf{q}_\alpha^2-\mu]\phi_{\mathbf{q}\alpha\sigma}+\sum_{\mathbf{q}\alpha}\left( R_{\mathbf{q}_\alpha}\phi_{\mathbf{q}\alpha\uparrow}^\dagger\phi_{\mathbf{q}\alpha\downarrow}+h.c.\right)\nonumber\\
   &+&\sum_{\mathbf{q}\alpha\beta\sigma}\left[\phi_{\mathbf{q}\alpha\sigma}^\dagger\phi_{\mathbf{q}\beta\sigma}\sum_{\alpha_1+\alpha=\alpha_2+\beta}\left(4g\phi_{0\alpha_1\sigma}^*\phi_{0\alpha_2\sigma}+2g_{\uparrow\downarrow}\phi_{0\alpha_1\bar{\sigma}}^*\phi_{0\alpha_2\bar{\sigma}}\right)+g\sum_{\alpha_1+\alpha_2=\alpha+\beta}(\phi_{0\alpha_1\sigma}\phi_{0\alpha_2\sigma}\phi_{\mathbf{q}\alpha\sigma}^\dagger\phi_{\mathbf{-q}\beta\sigma}^\dagger+h.c.)\right]\nonumber\\
   &+&\sum_{\mathbf{q}\alpha\beta}2g_{\uparrow\downarrow}\left[\sum_{\alpha_1+\alpha_2=\alpha+\beta}\left(\phi_{0\alpha_1\uparrow}^*\phi_{0\alpha_2\downarrow}^*\phi_{\mathbf{q}\alpha\uparrow}\phi_{\mathbf{-q}\beta\downarrow}+h.c.\right)+\sum_{\alpha_1+\alpha=\alpha_2+\beta}\left(\phi_{0\alpha_1\downarrow}^*\phi_{0\alpha_2\uparrow}\phi_{\mathbf{q}\alpha\uparrow}^\dagger\phi_{\mathbf{q}\beta\downarrow}+h.c.\right)\right],
\label{eq4}
\end{eqnarray}
\end{widetext}
where we have defined $\mathbf{q}_\alpha=\mathbf{q}+(2\alpha-1)\mathbf{K}$ and $R_\mathbf{q}=\lambda(q_x-iq_y)$.
To represent the gaussian action in a concise form, we shall define a column vector as $\Phi_\mathbf{q}=(\prod_\alpha\phi_{\mathbf{q}\alpha\uparrow}\phi_{\mathbf{q}\alpha\downarrow}\prod_\beta\phi_{\mathbf{-q}\beta\uparrow}^\dagger\phi_{\mathbf{-q}\beta\downarrow}^\dagger)^T$.
Then the gaussian action is written as $S_g=\frac{1}{2}\sum_{(\mathbf{q},iw_n)}\Phi_\mathbf{q}^\dagger\mathcal{G}^{-1}\Phi_\mathbf{q}-\sum_{\mathbf{q}\alpha}\frac{\xi_{\mathbf{q}\alpha}}{2}$, where $w_n$'s are the bosonic Matsubara frequencies, and $\xi_{\mathbf{q}\alpha}=[\mathbf{q}+(2\alpha-1)\mathbf{K}]^2+(2g+g_{\uparrow\downarrow})n_0-\mu$.
The matrix elements of the inverse Green's function $\mathcal{G}_{\alpha\sigma;\alpha^\prime\sigma^\prime}^{-1}(\mathbf{q},iw_n)$ can be conveniently constructed from Eq.~($\ref{eq4}$)(for details, please see~\cite{SM18}).

\begin{figure}[t]
\includegraphics[width=1.0\columnwidth,height=0.6\columnwidth]{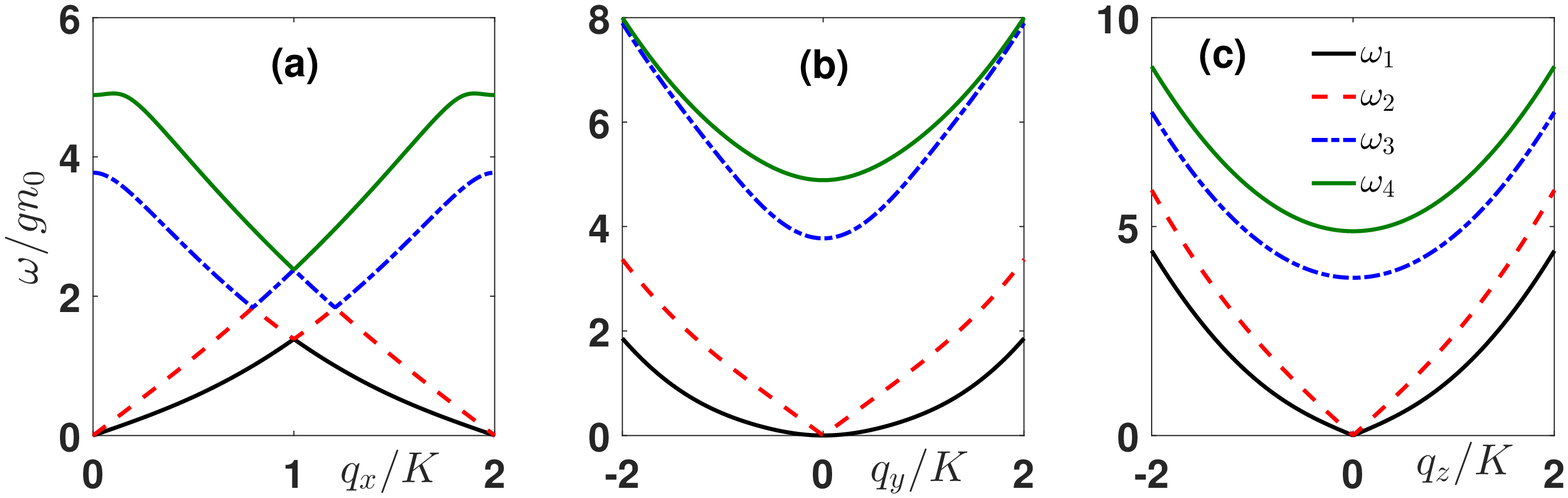}
\caption{(color online). The lowest four branches of quasi-particle excitation spectrum in the striped phase along (a) the x direction; (b) the y direction; and (c) the z direction.  There are two branches of gapless excitation: $\omega_1$ (black solid line) and  $\omega_2$ (red dash line), which correspond to two Goldstone modes resulting from breaking two continuous symmetries . The parameters used here are $\lambda/\sqrt{gn_0}=2$, $g_{\uparrow\downarrow}/g=2$. }
\label{fig2}
\end{figure}

The excitation spectrum of this system corresponds to the poles of the Green's function $\mathcal{G}$, and can be found by seeking solutions of the secular equation $det\mathcal{G}^{-1}(\mathbf{q},iw_n)=0$. The excitation predicted by Hamiltonian ($\ref{eq1}$) has been already calculated in the plane wave phase~\cite{BAY12,LIA13} where, despite the spinor nature of the system, it has only one gapless branch with rotonic structure except at the critical point $g_{\uparrow\downarrow}^c=g$ where the system enjoys a SU(2) rotation symmetry in pseudo-spin space.  We consider excitations propagating in three orthogonal directions (x, y, and z) and labeled with the wave vector $q_x$, $q_y$ and $q_z$, respectively. In Fig.~$\ref{fig2}$ we show the lowest four branches of excitation spectrum. The lowest two branches of excitation spectrum  are both gapless at the zero momentum $\mathbf{q}=\mathbf{0}$ and at the Brillouin wave vector $\mathbf{q}=2\mathbf{K}$. The peculiar feature which distinguishes the stripe phase from other uniform phases is the occurrence of double gapless bands, resulting from spontaneously-broken translational invariance symmetry and $U(1)$ gauge symmetry. As shown in panel (a), the excitation along the x direction displays a periodic structure in momentum space, and is fundamentally different from those in panel (b) and panel (c). In panel (b), the lowest branch shows a free particle-like behavior $\omega_1(0,\delta q_y,0)\propto (\delta q_y)^2$ along y direction, in stark contrast to the phonon-like behavior $\omega_1(0,0 ,\delta q_z)\propto \delta q_z$ in the z direction, shown in panel (c).

\begin{figure}
\includegraphics[width=1.0\columnwidth,height=0.6\columnwidth]{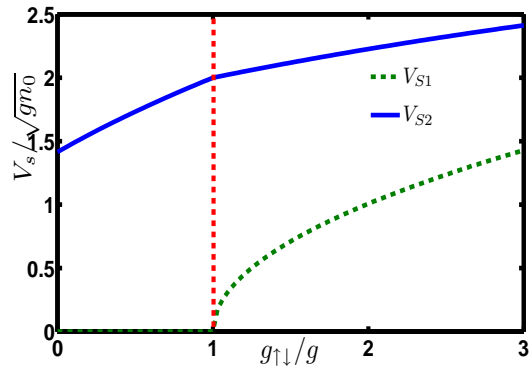}
\caption{(color online). The sound velocities $V_{S1}$ and $V_{S2}$ propagating along the x direction for the lowest two branches of excitation spectrum as a function of the interaction parameter $g_{\uparrow\downarrow}/g$. $V_{S1}$ stems from the broken translational invariance, and vanishes at the transition point $g_{\uparrow\downarrow}^{c}=g$ (marked by vertical red dash line) where the supersolid phase gives away to the plane wave phase. $V_{S2}$ is the conventional sound mode associated with spontaneously broken internal gauge symmetry. The parameters used here are: $\lambda/\sqrt{gn_0}=4$.}
\label{fig3}
\end{figure}

To identify which gapless band corresponds to the broken symmetry of translational invariance, we shall examine the behaviors of the sound velocities. The sound velocities propagating along in the x direction for the two gapless bands in the long wavelength limit are shown in Fig.~$\ref{fig3}$. The lower velocity $V_{S1}$ decreases monotonically as $g_{\uparrow\downarrow}/g$ decreases until it vanishes at the transition point where $g_{\uparrow\downarrow}^c/g=1$. This suggests that $V_{S1}$ corresponds to the Goldstone mode associated with spontaneously-broken continuous translational symmetry. The higher velocity $V_{S2}$ varies continuously as the interaction parameter $g_{\uparrow\downarrow}/g$ sweeps across the transition point. We judge that it is the conventional superfluid sound velocity stems from the U(1) gauge symmetry breaking.

\begin{figure}
\includegraphics[width=1.0\columnwidth,height=0.6\columnwidth]{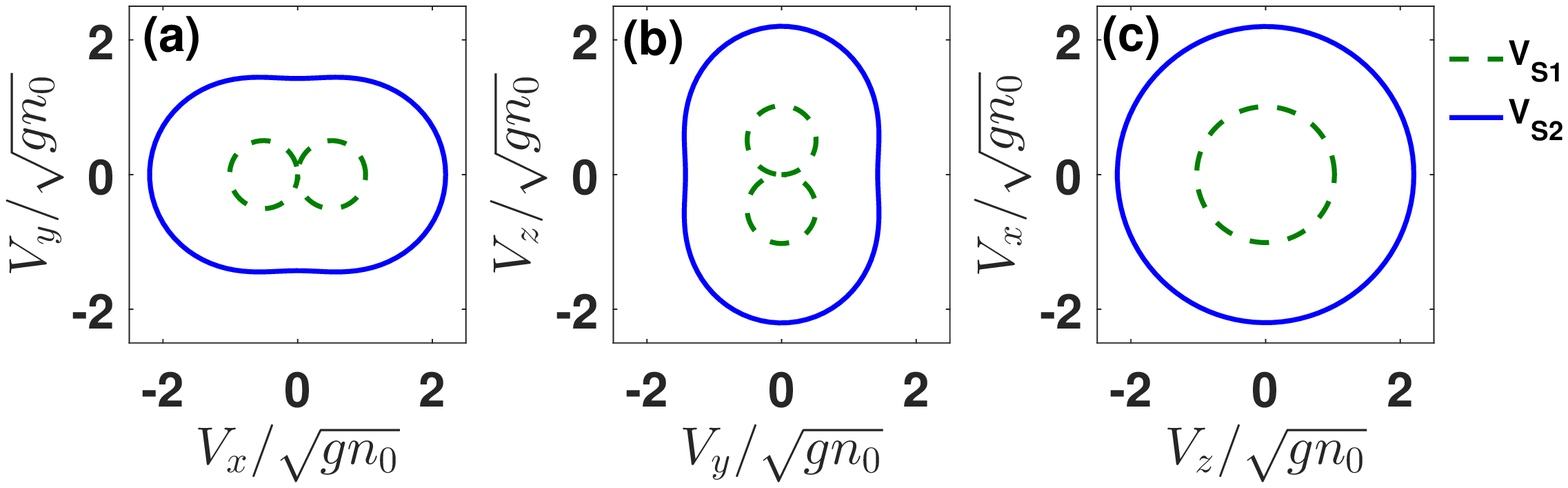}
\caption{(color online). Directional dependence of the sound velocities $V_{S1}$ and $V_{S2}$: (a) In the x-y plane; (b) In the y-z plane; and (c) in the z-x plane. Remarkably, the sound velocity of the lower branch $V_{S1}$ vanishes along y direction. The parameters used here: $g_{\uparrow\downarrow}/g=2$ and $\lambda/\sqrt{gn_0}=2$.}
\label{fig6}
\end{figure}

The directional dependence of the two branches of sound velocities is shown in Fig.~$\ref{fig6}$. Evidently both sound velocities enjoy the mirror symmetry of $V_S(\theta,\varphi)=V_S(\pi-\theta,\varphi)=V_S(\theta,\pi+\varphi)$. It is remarkable that the sound velocity of the lower branch $V_{S1}$ vanishes in the y direction, in stark contrast to that of the upper branch $V_{S2}$ which is always finite in all directions. Close inspection indicates that the sound velocities are slightly different in the x and z directions, reflecting an anisotropic nature induced by Rashba SOC.

\begin{figure}[t]
\includegraphics[width=1.0\columnwidth,height=0.6\columnwidth]{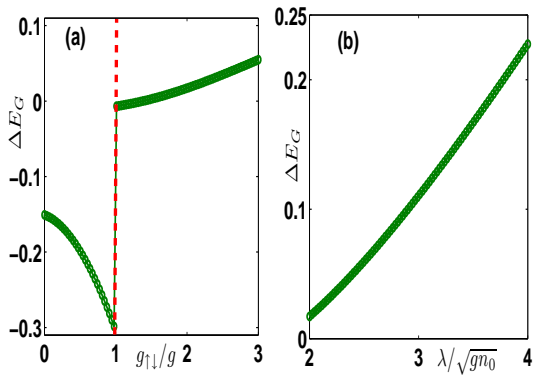}
\caption{(color online). Quantum fluctuation correction to the ground-state energy $\Delta E_G$ [in units of $V(gn_0)^{5/2}$] as a function of (a) the interaction parameter $g_{\uparrow\downarrow}/g$ with  $\lambda/\sqrt{gn_0}=2$ and (b) the spin-orbit coupling strength $\lambda/\sqrt{gn_0}$ with $g_{\uparrow\downarrow}/g=2$. At the transition point $g_{\uparrow\downarrow}^c$, the shift of the ground state energy is discontinuous, indicating that the transition is of a first order one involving with different order parameter symmetry. At the transition point, the plane-wave phase is preferred as its energy is lower than that of the supersolid phase. }
\label{fig4}
\end{figure}

The thermodynamic potential is given by $\Omega=-T\ln{\mathcal{Z}}=\Omega_0+\Omega_g=TS_0+\frac{T}{2}Tr\ln{\mathcal{G}^{-1}}-\frac{1}{2}\sum_{\mathbf{q}\alpha}\xi_{\mathbf{q}\alpha}$. It should be pointed out that the interaction parameters $g$ and $g_{\uparrow\downarrow}$ contained in $S_0$ should be renormalized to ensure non-divergent behaviors~\cite{STO09,AND04,REG18}. The quantum fluctuation correction to the ground state energy $\Delta E_G=\Omega_g$ is shown in Fig.~$\ref{fig4}$. As shown in panel (a), the sign of $\Delta E_G$ is reversed as the interspecies interaction strength $g_{\uparrow\downarrow}$ is tuned across the intraspecies interaction strength $g$. In the plane wave phase, $\Delta E_G$ decreases monotonically as $g_{\uparrow\downarrow}$ increases; while in the stripe phase, the trend is reversed. The discontinuity of the shift of the ground state energy at the transition point indicates that the phase transition is of first order. As shown in panel (b), spin-orbit coupling enhances the correction to the ground state energy. At the transition point, we conclude that the plane wave phase is favored over the stripe phase since the correction of the energy is lower, a direct verification of ``order from disorder" mechanism~\cite{WU11,HU12}.

\begin{figure}[t]
\includegraphics[width=1.0\columnwidth,height=0.6\columnwidth]{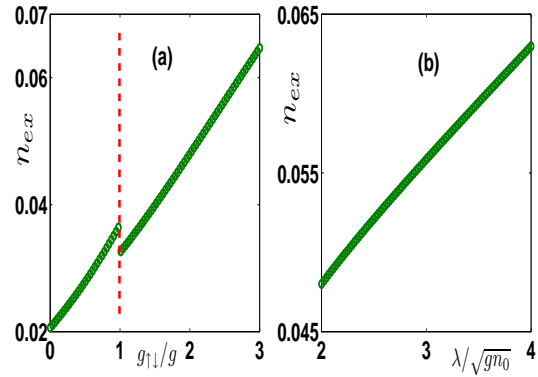}
\caption{(color online). The number density of excited particles $n_{ex}$ [in units of $(gn_0)^{3/2}$] due to quantum fluctuation as a function of (a) the interaction parameter $g_{\uparrow\downarrow}/g$ and (b) the spin-orbit coupling strength $\lambda/\sqrt{gn_0}$. The phase transition point is marked with a vertical red dash line. The parameters used here are: $\lambda/\sqrt{gn_0}=2$.}
\label{fig5}
\end{figure}

The type of symmetry that is broken has fundamental consequences on the system. Whereas a discrete symmetry results in robust states with gapped excitations, a continuous symmetry leads to an infinite number of degenerate ground states that can evolve from one to another without energy cost, making the system highly susceptible to fluctuations. For Bose gases with Rashba SOC, the quantum fluctuation will be greatly enhanced due to increased low-energy density of states~\cite{BAY12,ZHO13}. For the system to be stable we require that the quantum depletion should be finite. We evaluate the density of the excited particles due to quantum fluctuation via the Green's function:  $n_{ex}=\sum_{(\mathbf{q},iw_n)}\sum_{\alpha\sigma}\mathcal{G}_{\alpha\sigma,\alpha\sigma}(\mathbf{q},iw_n)$. By analyzing low-energy asymptotic behavior of the excitation spectrum, we have verified that there is no infra divergence, which renders that $n_{ex}$ is a finite quantity.  The number density of excited particles is shown in Fig.~$\ref{fig5}$. Clearly, both interspecies coupling and spin-orbit coupling enhance the quantum depletion. As seen in panel (a), at the transition point $g_{\uparrow\downarrow}=g$, $n_{ex}$ is discontinuous, characteristic of the first order phase transition.


So far, we have only concentrated on the zero-temperature properties. Finite-temperature effects will have vital effects on our system due to increased low-energy density of states due to SOC~\cite{BAY13,LIA14}. However, the celebrated Bogoliubov theory is only strictly valid at zero temperature. To extend it to a finite temperature where excitations proliferate, one needs to take account of interactions between excitations. At finite temperature, the Hartree-Fock-Bogoliubov approximation gives a gapped spectrum~\cite{GRI96}, violating the Hugenholz-Pines theorem~\cite{HUG59} and the Goldstone theorem. In this work, we shall undertake the Popov approximation~\cite{POP01}, which yields a gapless spectrum and provides a good description of Bose gases at finite temperatures.  Under the Popov approximation where anomalous average are neglected, the terms with three and four fluctuating fields in the action are approximated as follows~\cite{AND04}: $\phi_\sigma^\dagger\phi_\sigma\phi_\sigma\approx2{<}\phi_\sigma^\dagger\phi_\sigma{>}\phi_\sigma$, $(\phi_\sigma^\dagger\phi_\sigma)^2\approx4{<}\phi_\sigma^\dagger\phi_\sigma{>}\phi_\sigma^\dagger\phi_\sigma$ and $\phi_\uparrow^\dagger\phi_\uparrow\phi_\downarrow^\dagger\phi_\downarrow\approx{<}\phi_\uparrow^\dagger\phi_\uparrow{>}\phi_\downarrow^\dagger\phi_\downarrow+{<}\phi_\downarrow^\dagger\phi_\downarrow{>}\phi_\uparrow^\dagger\phi_\uparrow$. The shift of chemical potential can be obtained by requiring that the linear term in fluctuating fields vanishes: $\mu(T)=\mu(0)+(2g+g_{\uparrow\downarrow})n_{ex}$, with $n_{ex}=\sum_\sigma{<}\phi_\sigma^\dagger\phi_\sigma{>}$ being the density excited out of the condensates. The new gaussian action remains the same form as the original one, except that now $n_0$ becomes temperature dependent. In this way both $n_0(T)$ and $n_{ex}(T)$ can be determined self-consistently by fixing the total density  $n=n_0+n_{ex}$.

To sum up, we have shown that the supersolid phase exists in ultracold atomic condensates with Rashba spin-orbit coupling. The density distribution shows characteristic periodic modulation with a spatial period spontaneously chosen through translational symmetry breaking. Its elementary excitation propagating along the direction perpendicular to the stripe features double gapless bands. Quantum fluctuation correction to the ground state energy shows discontinuity at the phase transition suggests that it is of a first order one. Both interspecies coupling and spin-orbit coupling enhance quantum depletion of condensates. Our predictions bear consequences for experimental observation. The excitation spectrum and sound velocity can be probed by Bragg spectroscopy~\cite{CHE15}. An interesting extension to our work would be to map out the finite temperature phase diagram, which can be accessed in experiments~\cite{CHE14}. Experimental verification of our work is expected to contribute to a better understanding of supersolidity and emerging phenomena associated with breaking of  continuous symmetries.


 We acknowledge stimulating discussions with Lin Wen. This work was supported by the NSFC under Grants No. $11674058$.

\bibliographystyle{apsrev4-1}

\begin{thebibliography}{10}
\bibitem{REG18}%
  \BibitemOpen
  \ \bibinfo {note} {the thermodynamic potential posseses an ultraviolet
  divergence, an artifact of zero range interactions. It can be removed either
  by replacing the bare interaction parameters $g$ and $g_{\uparrow\downarrow}$
  with associated T-matrix or by substracting counter terms.}%
  \bibAnnoteFile{Stop}{REG18}%
\end{thebibliography}
%

\end{document}